\documentclass[published]{JHEP3}
\JHEP{11(2003)011}

\usepackage{epsfig}

\newcommand{\SU}{\mathop{\rm SU}\nolimits}
\newcommand{\tr}{\mathop{\rm tr}\nolimits}
\newcommand{\Tr}{\mathop{\rm Tr}\nolimits}

\title{On plane wave and vortex-like solutions of noncommutative  Maxwell-Chern-Simons theory}

\author{Garnik Alexanian and Manu~B.~Paranjape\\
        Groupe de physique des particules, Universit\'e de Montr\'eal\\
	C.P.\ 6128, succ.\ centre-ville, Montr\'eal, Qu\'ebec, Canada H3C 3J7\\
	E-mail: \email{garnik@lps.umontreal.ca},
	        \email{paranj@lps.umontreal.ca}}

\author{Daniel Arnaudon\\
        Laboratoire d'Annecy-le-Vieux de Physique Th\'eorique LAPTH\\ 
	CNRS, UMR 5108, Universit\'e de Savoie\\
	B.P.~110, F-74941 Annecy-le-Vieux Cedex, France\\
	E-mail: \email{daniel.arnaudon@lapp.in2p3.fr}}

\abstract{We investigate the spectrum of the gauge theory with
  Chern-Simons term on the noncommutative plane, a modification of the
  description of the Quantum Hall fluid recently proposed by Susskind.
  We find a series of the noncommutative massive ``plane wave''
  solutions with polarization dependent on the magnitude of the
  wave-vector.  The mass of each branch is fixed by the quantization
  condition imposed on the coefficient of the noncommutative
  Chern-Simons term.  For the radially symmetric ansatz a vortex-like
  solution is found and investigated. We derive a nonlinear difference
  equation describing these solutions and we find their asymptotic
  form.  These excitations should be relevant in describing the
  Quantum Hall transitions between plateaus and the end transition to
  the Hall Insulator.}

\received{October 13, 2003}
\accepted{November 7, 2003}

\begin{document}

\section{Introduction}\label{section1}

Since their reappearance as the effective low-energy descriptions of
some string theory configurations~\cite{Snyder:1946qz} noncommutative
(NC) field theories have been a subject of much
attention~\cite{Douglas:2001ba}.  Among the flurry of activities
devoted to their study there was an interesting proposal by
Susskind~\cite{Susskind:2001fb} on the possible relation between NC
Chern-Simons theory and the Quantum Hall (QH) effect. Following this
paper there was a considerable interest in the NC versions of the
Chern-Simons (CS) theory and its
properties~\cite{Polychronakos:2000zm}--\cite{Fradkin:2002qw}.  There
are many indications that noncommutativity can indeed be an
interesting mathematical tool for describing 2d electron gases in
strong magnetic fields~\cite{Fradkin:2002qw}.  It would be interesting
if the NC description would lead to some specific experimentally
observable differentiating phenomena.

One of the most useful tools for studying the low energy properties of
the QH systems are effective lagrangians.  It has been known for some
time that depending on location on the quantum Hall conductivity graph
the effective charge carriers differ between different plateaus and
transitions on the transverse conductivity plot $\sigma_{xy}$,
including the transition to the Hall insulator.  It has been argued
that these carriers are ``particles'' and ``vortices'' on either side
of the transition and there exists an effective lagrangian describing
each type of the carrier (see~\cite{Burgess:2000kj} and references
therein).  In two space dimensions in the presence of a perpendicular
magnetic field there is a unique gauge-invariant lagrangian which is
dominant in the long-wavelength limit --- the Chern-Simons action. Due
to this property one can assume that whatever the effective actions
for particles or vortices are, they must have similar functional form
in the long-wave limit, given by that of the CS. A pure CS theory can
only describe the plateaus, it cannot support longitudinal
conductivity. Hence as a first correction, the Maxwell term (i.e.\ $F^2$) is
added, allowing for longitudinal conductivity and the possibility to
describe the transition regions.  It was observed
by~\cite{Burgess:2000kj} that an interesting symmetry of the QH
systems --- the ``modular group'' symmetry of the complex conductivity
plots could be derived just from this similarity of their effective
actions for the two types of charge carriers --- and it is true even
beyond the linear approximation~\cite{Burgess:2001sy}.  For our
purposes, the essential point is that one cannot restrict oneself to
just the lowest order term. In order to study the dynamics of charge
carriers through the transition regions it is important to include the
Maxwell term as well.  In the effective action the coefficient of the
Maxwell term is just $\sigma_{xx}$, which is necessary for the action
of the modular group~\cite{Burgess:2000kj}.

This brings us to a motivating point of our paper: if there is an
effective NC theory describing the QH system, what would the
noncommutativity imply for the duality?  Do the effects of the order
of $\theta$ (NC parameter which is $\sim 1/\rho$, $\rho$ --- electron
density in Susskinds proposal) spoil the resulting ``modular group''
action?  If so, would the effect be detectable?  In order to start
answering these questions one first needs to find what the effective
``charge carriers'' for the theory in the different phases are. It
therefore calls for investigation of the classical solutions of the NC
Maxwell-CS theory.

In what follows we will study possible classical solutions of this
action. In the next section~\ref{section2} we give a short review of
the NC Maxwell-CS theory and derive equations of motion for the model.
In the section~\ref{section3} we will find the exact solutions
corresponding to the plane waves with ``massive'' dispersion relation.
In the section~\ref{section4} we will derive the nonlinear difference
equation describing the static configurations that possess a non-zero
vorticity and in general look like ``dyons'' (in the sense that there
is ``magnetic flux'' and electric charge attached to them,
``magnetic'' and ``electric'' being just commutators of $[D_i,D_j]$
and $[D_0,D_i]$).  We will derive some general properties of these
solutions and construct some of them numerically.  Some conclusions
and proposals for further research are given in the end.

\section{Noncommutative Chern-Simons theory}\label{section2}

Before making any connection with the Quantum Hall effect, 
let us first consider noncommutative versions of the
Yang-Mills ---  Chern-Simons theory.  
In this chapter we will closely follow~\cite{Nair:2001rt,Polychronakos:2000zm}.
We consider the the three-dimensional noncommutative
space with coordinates $x^\mu = (x^0,x^1,x^2)$
with 
\begin{eqnarray}
x^0&=&t\,,\qquad [x^1,x^2]=\,i\,\theta\,, \qquad \theta > 0 
\nonumber \\{}
[ t,x^{i} ]&=&0\,,\qquad i=(1,2)
\label{CommRell}
\end{eqnarray} 
which can be written as
\begin{eqnarray}
[x^\mu,x^\nu]&=&i\,\theta^{\mu\nu}
\nonumber\\
\theta^{12}&=&-\theta^{21}=\theta\,,\qquad
\theta^{01}=\theta^{10}=0\,.
\end{eqnarray}
The noncommutative analog of the integral over all three coordinates
in space time gets replaced by the mixed expression involving integral
over time but trace over Fock space of the harmonic oscillator
representing the spatial integral in the $(x^1,x^2)$ plane. Following
conventions of~\cite{Nair:2001rt}, we use normalization as
\begin{equation}
\int\,d^3\,x\,\rightarrow\,\int\,dt\,(2\pi\theta)\,\tr\,.
\end{equation}

One way to introduce this noncommutativity in any 2D theory described
by some action $S$ is through the so called star ($''\star''$)
product.  One can show~\cite{Groenewold:kp} that the effect of having
noncommuting coordinates can be modeled by changing the multiplication
rule between functions on space time. For any two functions $\phi(x)$
and $f(x)$ one has
\begin{equation}
\tr\left(f({\hat x})\,\phi({\hat x})\right)\,\sim\, \int\,d^2x\,
f({x})\,e^{\frac{i}{2}\,{\overleftarrow{\partial}}_i\theta^{ij}
  {\overrightarrow{\partial}}_j}\,\phi({x}) \,\equiv\,\int d^2x\,
f(x)\star\phi(x)\,,
\end{equation}
where the first derivative in the exponent acts to the left and second
one to the right. This product is obviously noncommutative and
introduces additional nonlocality in the theory with interactions
(i.e.\ having terms of order three or more in the action).  This is
because while
\begin{eqnarray} 
\int f(x)\star\phi(x)\,d^2x&=&\int f(x)\phi(x)\,d^2x\,,\qquad
{\rm but}
\nonumber\\ 
\int f(x)\star\phi(x)\star\xi(x)\,d^2x &\not=&
\int f(x)\phi(x)\xi(x)\,d^2x\,.
\end{eqnarray}
And therefore any action which has a term of order three or more will
contain infinite number of derivatives, making the theory essentially
nonlocal.

Using this formalism, it is easy to write the NC version of the CS
theory
\begin{equation}
S=\,\frac{1}{4\pi\nu}\,
\int\,dt\,d^2x\,\left(A_\mu(x)\partial_\nu A_\gamma(x)+
i\frac{2}{3}A_\mu(x)\star A_\nu(x)\star A_\gamma(x)\right)
\epsilon^{\mu\nu\gamma}\,.
\end{equation} 
Notice that in the commutative limit $\theta\rightarrow 0$ the
non-abelian term disappears and this expression becomes the usual
abelian CS action.

While this ``star-product'' approach proves to be very convenient in
discussing the perturbation theory in this new, non-commutative field
theories (see for example~\cite{Guralnik:2001ax,Aref'eva:1999sn}), it
unfortunately hides the discreteness of this space-time picture. One
essentially trades-off the Hilbert space picture of the underlying
space-time to the non-locality of the action. Therefore, we will not
use it in the following text, but will follow approach
of~\cite{Nair:2001rt,Polychronakos:2000zm} and work directly with
operators rather then with functions.

Firstly, we notice that in the NC theory described by the
commutation relations~(\ref{CommRell}) the \emph{derivatives} with
respect to the coordinates $x^1,x^2$ have to be represented as
operators as well:
\begin{eqnarray}
&{\rm Since}&\qquad [\partial_{i},x^j]=\delta_{ij}\,,
\qquad {\rm and} \qquad [x^1,x^2]=i\,\theta 
\nonumber\\
&{\rm then}&\qquad \partial_{1}=\frac{ix^2}{\theta}\,,\qquad
\partial_{2}=\frac{-ix^1}{\theta}\,,\qquad
[\partial_1,\partial_2]=-\frac{i}{\theta}\,.
\end{eqnarray}
Following~\cite{Nair:2001rt,Polychronakos:2000zm}, we will define the
hermitean covariant derivative operators as
\begin{equation}
D_\mu=-i\partial_\mu+A_\mu(x)\,,
\end{equation}
where $A_\mu(x)$is a generic hermitean operator representing a gauge
field.  Let us first write the NC CS action in these variables.
\begin{equation}
S = 2\pi \theta \int dt \tr \left( i{2\over 3} D_\mu D_\nu D_\alpha
-\omega_{\mu\nu} D_\alpha \right)\epsilon^{\mu\nu\alpha}\,.
\label{CS1}
\end{equation}
Here $\omega_{\mu\nu}$ is an antisymmetric tensor defined as follows
(its spatial part is inverse of $\theta^{\mu\nu}$)
\begin{equation}
\omega_{12}=-\omega_{21}=-\frac{1}{\theta}, \qquad
\omega_{01}=\omega_{02}=0\,.
\end{equation}
The cubic term in this action will produce (upon expansion in
$A_\mu$'s and $\partial_\nu$) the term proportional to $A^3$, the
quadratic term $\sim A\,\partial\,A$ and a linear term, given by
\begin{equation}
\sim\, -2i\,\partial_i\,\partial_j\,D_0\epsilon^{ij0}\,
=-2i\,[\partial_1,\partial_2]\,D_0 = -\frac{2}{\theta}\,D_0\,.
\end{equation}
This last term is, of course, the ``one dimensional'' CS term and is
absent in the commutative theory ($\sim 1/\theta$). It has to be
subtracted precisely by the linear term of the equation~(\ref{CS1}).
This condition also fixes the relative sign between these two terms.
For our purposes it is more convenient to introduce holomorphic and
anti-holomorphic combinations as
\begin{equation}
D=\frac{D_1+i\,D_2}{\sqrt{2}}\qquad
{\bar D}=\frac{D_1-i\,D_2}{\sqrt{2}i}\,.
\end{equation}
Since $D_\mu$'s are hermitean ${\bar D}=D^\dagger$.  With this
definition of $D,\bar D$ we get
\begin{equation}
S = \lambda\, 2\pi \theta \int dt \tr\left( -2\,[D,\bar{D}] D_0 +
\frac{2}{\theta} D_0 \right).
\end{equation}
In order to make the following calculations easier, it is convenient
to switch to the dimensionless version of operators $D,\bar D$ and
$D_0$ by re-scaling them as
\begin{equation}\
D\,\rightarrow\,\frac{D}{\sqrt{\theta}}\,,\qquad {\bar
  D}\,\rightarrow\,\frac{{\bar D}}{\sqrt{\theta}} \qquad
D_0\,\rightarrow\,\frac{D_0}{\sqrt{\theta}}\,.
\label{scale}
\end{equation}
The re-scaling of $D_0$ also implies that $dt\rightarrow \sqrt{\theta
  d}{\tilde t}$ and new ``time'' is dimension-less as well.
Therefore, the final form of the CS action that we are going to work
with reads
\begin{equation}
S = \frac{2\,\lambda}{\theta^{\frac{3}{2}}}\, (2\pi
\theta^{\frac{3}{2}}) \int d{\tilde t} \tr\left( -[D,\bar{D}] + 1
\right)\,D_0\,.
\label{fCS}
\end{equation}

Now let us briefly turn to the gauge theory action.  In terms of the
$D,{\bar D}$ and $D_0$ operators before re-scaling the standard YM
action is
\begin{eqnarray}
S_{\rm YM} &=& \int dtd^2x\,\frac{1}{4g^2}\, F_{\mu\nu}\,F^{\mu\nu}=
\int dt\,d^2x\,\frac{1}{4g^2}\,[D_\mu,D_\nu][D^\mu,D^\nu]
\nonumber\\
&=&\int dt\,d^2x\,\frac{1}{4g^2}
\left(-2[D_0,D_i][D_0,D_i]+[D_i,D_j][D_i,D_j]\right)
\nonumber\\
&=&(2\pi\theta)\int dt\tr\frac{1}{4g^2}
\left(-4[D_0,D][D_0,{\bar D}]-2[D,{\bar D}][D,{\bar D}]\right).
\end{eqnarray}
After performing (\ref{scale}) the final form on the action is
\begin{equation}
S_{\rm YM} =(2\pi\theta^{\frac{3}{2}}) \int d{\tilde
  t}\tr\frac{-2}{4g^2\,\theta^2} \left(2[D_0,D][D_0,{\bar D}]+[D,{\bar
    D}][D,{\bar D}]\right).
\label{fYM}
\end{equation}
We kept the factor $(2\pi\theta^{{3}/{2}})$ separate from the rest
of the expressions~(\ref{fCS}) and~(\ref{fYM}) since it does not
contribute to the relative coefficient between two different actions
which will become important later.  Let us now consider the total
expression
$$
(2\pi\theta^{\frac{3}{2}})\int d{\tilde t}\tr\left\{
\,\frac{-2}{4g^2\,\theta^2} \left(2[D_0,D][D_0,{\bar D}]+[D,{\bar
    D}][D,{\bar D}]\right)+ \frac{2\,\lambda}{\theta^{3/2}}\, \left(
-[D,\bar{D}] + 1 \right)\,D_0\right\}.
$$
What is important is just the relative factor between the two actions,
so it makes sense to express this as
\begin{equation}
\frac{(2\pi\theta^{3/2})}{2g^2\,\theta^2}\int_{\tilde t}\tr\left\{
\left(-2[D_0,D][D_0,{\bar D}]-[D,{\bar D}][D,{\bar D}]\right)+
     {2\tilde\lambda}\, \left( -[D,\bar{D}] + 1 \right)\,D_0\right\},
\label{fEQ}
\end{equation}
where 
\begin{equation}
\tilde\lambda=2\,g^2\,\theta^{1/2}\,\lambda\,.
\label{lambda}
\end{equation}
This is the final form of the action we are going to work with.  It
what follows, we can treat $D, \bar D$ and $D_0$ as independent variables,
since in the non-commutative space gauge fields are operators in the
same Hilbert space as the original $\partial_i$'s.
Varying~(\ref{fEQ}) with respect to $D_0$ one gets
\begin{equation}
-[D,[D_0,{\bar D}]]-[{\bar D},[D_0,D]]={\tilde\lambda}([D,{\bar D}]-1)\,.
\end{equation}
In turn, for the $\delta {\bar D}$ variation the equation is
\begin{equation}
[D_0,[D_0,D]]+[D,[D,\bar D]]={\tilde\lambda}[D_0,D]\,.
\end{equation}

It is well known that in the commutative version of the CS action for
the non-abelian gauge field the overall coefficient on the action must
be quantized due to the topology of the $\SU(N)$. It so happens that
a similar effect takes place in the noncommutative version as well,
where the effective non-abelian nature of the fields is due to the space
noncommutativity.  As it was shown in~\cite{Nair:2001rt}, the
condition is:
\begin{equation}
4\pi\lambda=n\,, \qquad  n\in Z-\{0\}\,.
\end{equation}

\section{Noncommutative plane waves}\label{section3}

Let us now describe the plane wave solutions of the theory.  Generic
form of the equations of motion is given by
\begin{eqnarray}
-{\ddot{D}} + [D[D,D^\dagger]]&=& -i \,\lambda\,\dot{D}
\label{eq0}\\
i\,[D,{\dot{D}}^\dagger]+i\,[D^\dagger,\dot{D}]&=& 
\lambda[D,D^\dagger]-\lambda 
\label{eq1}
\end{eqnarray}
(to simplify notation we use $\lambda$ instead of $\tilde\lambda$ in
what follows).  The ``vacuum'' solution is given by $D=a$,
$D^\dagger=a^\dagger$, where $[a,a^\dagger]=1$ --- standard creation and
annihilation operators for the harmonic oscillator.  Let us look for
the solution of the equations~(\ref{eq0}) and~(\ref{eq1}) using the
following ansatz:
\begin{equation}
D=a+ e^{i\,\omega\,t}R_1\left(k\,a+k^*\,a^\dagger\right)+
e^{-i\,\omega\,t}R_2\left(k\,a+k^*\,a^\dagger\right),
\label{anzats}
\end{equation}
where $k, k^*$ are complex numbers and $R_1(x)$ and $R_2(x)$ are some
(possibly complex) functions of the argument that can be expanded in
series.  Using the property that
\begin{equation}
\left[R_i\left(k\,a+k^*\,a^\dagger\right),a^\dagger\right]= \left[x, a^\dagger\right]\,
\left.\frac{\partial R_i(x)}{\partial
  x}\right|_{x=k\,a+k^*\,a^\dagger}\!=\,\,
k\,R_i^{'}(k\,a+k^*\,a^\dagger)
\end{equation}
(which is valid only if $k$ and $k^*$ are c-numbers) the
equation~(\ref{eq0}) can be brought to the following form:
\begin{eqnarray}
\omega^2\,R_1+(k^*)^2\,{R^*}^{''}_2+{|k|}^2\,R_1^{''}&=&
\omega\,\lambda\,R_1
\nonumber\\
\omega^2\,R_2+(k^*)^2\,{R^*}^{''}_1+{|k|}^2\,R_2^{''}&=&
-\omega\,\lambda\,R_2\,.
\label{system0}
\end{eqnarray}
At this point it is possible to assume that
\begin{equation}
R_1^{''}=-R_1\,,\qquad R_2^{''}=-R_2\,,\qquad R_{1,2}\sim
A_{1,2}\,e^{ix}+B_{1,2}e^{-ix}
\end{equation}
(an interesting case where $R^{''}_{1,2}=+R_{1,2}$ will be considered
later).  This reduces the system to
\begin{eqnarray}
\omega^2\,R_1-(k^*)^2\,{R^*}_2-{|k|}^2\,R_1&=&
\omega\,\lambda\,R_1
\nonumber\\
\omega^2\,R_2-(k^*)^2\,{R^*}_1-{|k|}^2\,R_2&=&
-\omega\,\lambda\,R_2\,.
\label{system1}
\end{eqnarray}
Taking the complex conjugate of the second equation the resulting system
can be then written in matrix form
\begin{equation}
\pmatrix{\omega^2-{|k|}^2-\omega\,\lambda & -(k^*)^2 \cr -(k)^2 &
  \omega^2-{|k|}^2+\omega\,\lambda} \pmatrix{R_1 \cr {R^*}_2}= 0\,.
\label{matrix}
\end{equation}
The existence of the nontrivial solution for $R_i$ will require that
determinant of the matrix on the left-hand side of~(\ref{matrix}) is
zero, which in turn produces the following dispersion relation
\begin{equation}
\omega^2(\omega^2-\lambda^2-2|k|^2)=0\,.
\end{equation}
The solution $\omega=0$ is actually trivial --- it is just a gauge
transformation. Indeed, for this branch one has that
\begin{eqnarray}
R_2&=&-\frac{k}{k^*}R_1^*\,, \qquad D=a+R_1\left(ka+k^*a^\dagger\right)
-\frac{k}{k^*}R_1^*\left(ka+k^*a^\dagger\right)
\nonumber\\
D&=&U^\dagger\,a\,U,\ \ U^\dagger=U^{-1}\,,\qquad U=\left. \exp\left(
{\frac{1}{k^*}\int R_1(x)-\frac{1}{k}\int R_1^*(x)} \right)
\right|_{x=ka+k^*a^\dagger}\,.
\end{eqnarray}

For the $\omega \ne 0$ solution one then has the ``massive'' modes,
given by
\begin{equation}
\omega=\pm\sqrt{\lambda^2+2|k|^2}\,.
\end{equation}
Finding the eigenvectors of the equation of the matrix
in~(\ref{matrix}) corresponding to these eigenvalues gives
\begin{eqnarray}
R^*_2&=&\left(\frac{\omega-\lambda}{\omega+\lambda}\right)
\,e^{2i\,\phi}\,R_1 
\nonumber \\
k&=&|k|\,e^{i\,\phi}\,, \qquad \omega^2=\lambda^2+2|k|^2\,.
\label{wave0}
\end{eqnarray}
However, this solution only satisfies equation~(\ref{eq0}) so far. One
has to check that it is going to be compatible with
equation~(\ref{eq1}) as well.  Quite remarkably, upon substituting
ansatz~(\ref{anzats}), this equation reduces to the derivative of
condition~(\ref{wave0}).
\begin{equation}
R^{*'}_2=\left(\frac{\omega-\lambda}{\omega+\lambda}\right)
\,e^{2i\,\phi}\,R_1^{'}\,.
\end{equation}
Thus~(\ref{wave0}) is a solution. While it is not surprising that the
dispersion relation is massive (the same as in the commutative
YM-CS theory~\cite{Deser:vy}), it shows some interesting properties ---
unlike the usual (i.e.\ commutative) plane waves there is a dependence
of the polarization on the magnitude of the wave vector, $k$
(see~\cite{Guralnik:2001ax} for the noncommutative QED case).  In the
commutative but non-abelian case the plane wave solutions will only
exist for a fixed direction in the color space.  Noncommutativity, a
space-time feature here, will make the abelian theory effectively
non-abelian, thus translating this property from the ``color'' space
to the polarization of the waves.

Finally, let us consider the $R^{''}_{1,2}=+R_{1,2}$ choice.  It
causes a sign change for all the $k^2$ terms in the equations starting
from~(\ref{system0}) and leads to the following expressions:
$$
\omega=\pm\sqrt{\lambda^2-2|k|^2}, \ k<{\lambda\over\sqrt{2}}\,,\qquad
R^*_2=-\left(\frac{\omega-\lambda}{\omega+\lambda}\right)
\,e^{2i\,\phi}\,R_1\,,\qquad R_{1,2}\sim A_{1,2}e^{x}+B_{1,2}e^{-x}\,.
$$ 
These solutions are not normalizable and represent excitations that
are ``glued'' to the boundary and presumably only make sense for the
finite size droplet.

\section{Static gauge solutions}\label{section4}

In this section we will look for vortex type solutions taking the
``static'' gauge ansatz, defined by the following condition:
\[
D_t=\phi(N)\,,
\]
where $N$ is the number operator, and $\phi$ is a well defined
function of $N$.  Such functions are essentially defined through the
spectral representation (see for example Reed and Simon).  All fields
are taken to be independent of the temporal coordinate, hence
$\partial_t$ effectively vanishes.  The corresponding ansatz for the
spatial part is given by
\[
D=f(N)a\qquad D^\dagger=a^\dagger f^\dagger (N)\,.
\] 
In the computations that follow we use the following identities which
can be established using the spectral representation, for example:
\begin{eqnarray}
[a,g(N)]&=&a(g(N)-g(N-1))=(g(N+1)-g(N))a
\nonumber\\{}
[a^\dagger ,g(N)]&=&-a^\dagger (g(N+1)-g(N))=(g(N)-g(N-1))a^\dagger\,.
\nonumber
\end{eqnarray}
Before proceeding to the equations of motion, there are two
expressions that are useful to compute.  One is the ``magnetic field''
$B(N)=[D,D^\dagger]$ while the other is $E= [D_t,D]$ which corresponds
to the holomorphic component of the ``electric'' field.  We find
\begin{equation}
B(N)=(N+1)f^\dagger(N)f(N)-Nf^\dagger(N-1)f(N-1)
\label{magnetic}
\end{equation}
while
\begin{equation}
E=-f(N)(\phi(N+1)-\phi(N))a\,.
\label{electric}
\end{equation}
If for the solution neither of these changes then it is just the gauge
transformation as discussed in the previous section ($\omega=0$ modes).

Now let us proceed to the equations~(\ref{eq0}) and~(\ref{eq1}) in
terms of these variables.  The Amp\`ere law becomes
\begin{equation}
\left[D_t,\left[D_t,D\right]\right]+\left[D,\left[D,D^\dagger \right]\right]=\lambda[D_t,D]\label{ampere}
\end{equation}
while the Gauss law becomes
\begin{equation}
-\left[D^\dagger,\left[D_t,D\right]\right]-\left[D,\left[D_t,D^\dagger\right]\right]=\lambda(\left[D,D^\dagger\right]+1)\,.
\label{gauss}
\end{equation}
Using definition
$$
Q(N)=\phi(N)-\phi(N+1)
$$
we get the following operator equations for~(\ref{ampere})
and~(\ref{gauss}) the Amp\`ere law and the Gauss law, respectively,
\begin{eqnarray}
f(N)(Q(N))^2a+f(N)(B(N+1)-B(N))a&=&\lambda f(N)Q(N)a\qquad
\label{ampere1}\\
2\left((N+1)f^\dagger(N)f(N)Q(N)-Nf^\dagger(N-1)f(N-1)Q(N-1))\right)&=& \lambda (B(N)-1)\,.
\label{gauss1}
\end{eqnarray}
The Ampere law, equation~(\ref{ampere1}) can be simplified further by
multiplying on the right by $a^\dagger$.  This yields the operator
$N+1$ as a common, positive, right multiplying factor in each term,
which we can then remove, since it never vanishes.  Similarly the
function $f(N)$ appears as a left multiplying factor in each term.
With the hypothesis that $f(N)$ never vanishes, we can also remove it
giving the equation
\[
Q(N)^2+(N+2)f^\dagger(N+1)f(N+1)-2(N+1)f^\dagger(N)f(N)+Nf^\dagger(N-1)f(N-1)=\!-\lambda
Q(N)
\]
which can be written as
\begin{equation}
\frac{1}{4}(2Q(N)-\lambda)^2
+\nabla^2\left((N+1)f^\dagger(N)f(N)\right)=\frac{\lambda^2}{4}\,,
\label{ampere2}
\end{equation}
where $\nabla^2$ is the appropriate discrete laplacian operator.  The
Gauss law, equation~(\ref{gauss1}) can be re-written as
\[
(N+1)\left(2Q(N)-\lambda\right)f^\dagger(N)f(N)-N(2Q(N-1)-\lambda)f^\dagger(N-1)f(N-1)=-\lambda\,.
\]
This equation can actually be solved as an operator equation, easily
seen by using the spectral representation, to give
\begin{equation}
(2Q(N)-\lambda)f^\dagger(N)f(N)=-\lambda\,.
\label{gauss2}.
\end{equation}
Replacing for $(2Q(N)-\lambda)$ in equation~(\ref{ampere2}) yields
\begin{equation}
{1\over 4}\left({\lambda\over f^\dagger(N)f(N)}\right)^2+
\nabla^2\Big((N+1)f^\dagger(N)f(N)\Big) ={\lambda^2\over
  4}\,.
\label{ampere3}
\end{equation}
Defining $u(N)=(N+1)f^\dagger(N)f(N)$ and $g^2=\lambda^2/4$ we get the
nonlinear recurrence relation (in the spectral representation)
\begin{equation}
u_{n+1}-2u_{n}+u_{n-1} +g^2\left({n+1}\over
u_{n}\right)^2=g^2\,.\label{recurrence}
\end{equation}
$u_{-1}$ is taken to be zero as can be seen from consistency with its
definition.  Hence specifying the value of $u_0$ is sufficient to
determine all subsequent values of $u_n$ for $n\geq 1$.

Let us study the simplest solution of our system of
equations~(\ref{gauss2}) and~(\ref{ampere3}). One obvious possibility
is to take $Q(N)=0$ and $f^\dagger(N)f(N)=1$ since the (discrete)
laplacian of $N+1$ evidently vanishes.  This corresponds to the
``vacuum'' solution of the quantum Hall system of Susskind and its
gauge transformations.  The ``magnetic field'' given
by~(\ref{magnetic}) is $[D,D^\dagger]=1$ and the energy ${\cal
  E} \sim \Tr (B^2-1)=0$.  The quasi-hole/particle excitation that are
found in the pure Chern-Simons theory of Susskind are actually absent
in this theory. Indeed, it seems that the vacuum solution is unstable
due to the addition of the Yang-Mills term.

\FIGURE[t]{\epsfig{file=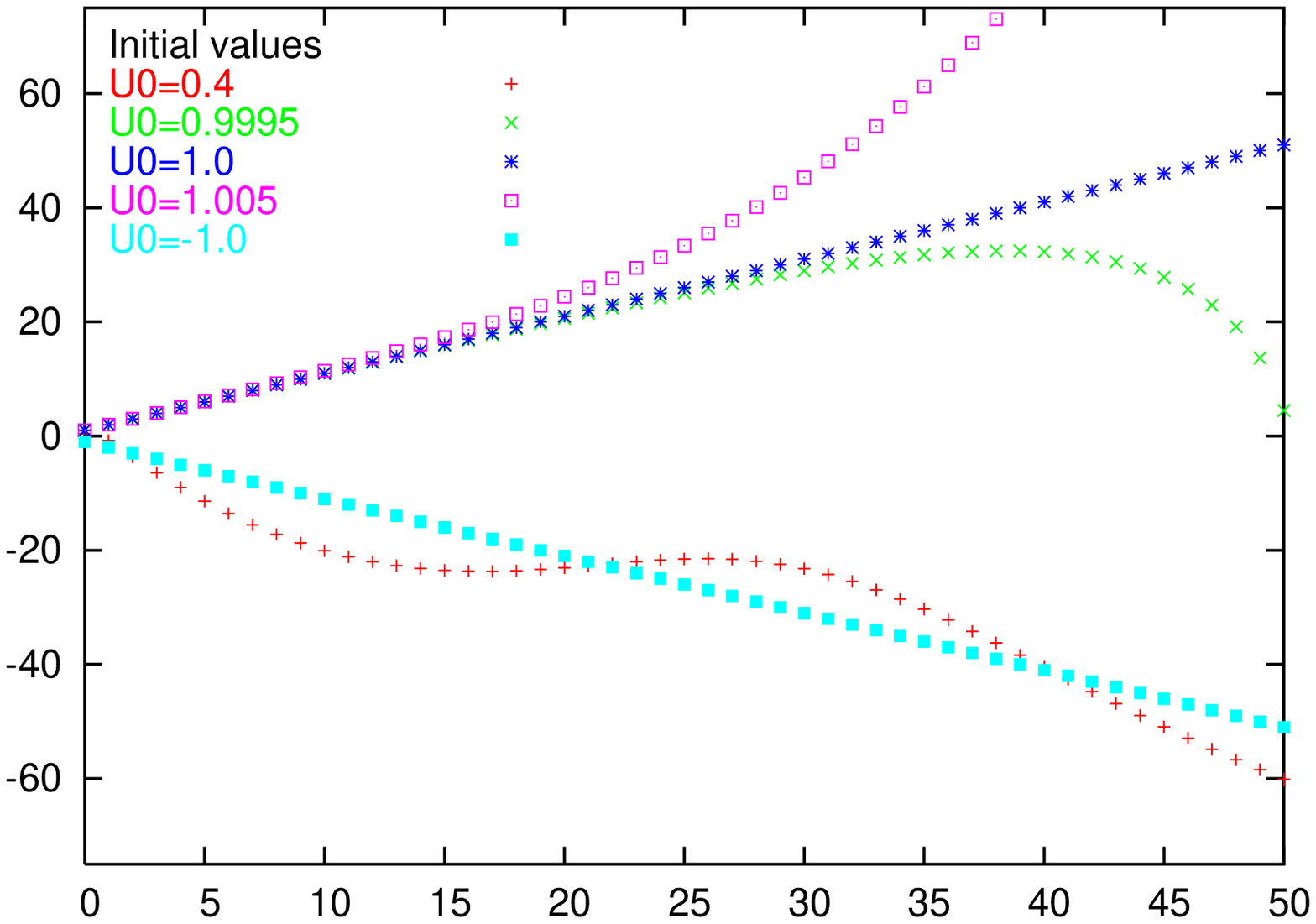,scale=0.6,angle=-90,clip=}%
\caption{Behavior of the solutions of the equation~(\ref{recurrence})
  for the different values of $u_0$.  The plot shows $u_n$ for $n$
  between 0 and 50.\label{fig1}}}

Numerical analysis of the recurrence relation~(\ref{recurrence})
yields the following intriguing behavior, some of which is shown on
figure~\ref{fig1}.  It turns out that the type of solution one gets greatly depends on
the initial value $u_0$.  If $u_0$ is taken to be one, we get the
above mentioned vacuum solution (data with $u_0=1.0$).  However if
$u_0$ is perturbed even slightly greater than one, the recurrence
relation is dominated by the right hand side and values for subsequent
$u_n$ become large (data with $u_0=1.005$). When this is the case, the
nonlinear term on the left hand side becomes negligible, yielding a
simple recurrence relation
\begin{equation}
u_{n+1}-2u_{n}+u_{n-1} =g^2
\label{recurrence1}
\end{equation}
with solution
\[
u_n\sim g^2\frac{n(n+1)}{2}\,.
\]
This solution is asymptotic to the solution found
by~\cite{Khare:2001ee} for the Yang-Mills-Chern-Simons theory with the
pure $D^3$ expression for the Chern-Simons term.  If $u_0$ is
perturbed slightly less than one, the behavior is very interesting but
always pathological.  It can essentially only be analyzed numerically
and we find that it is possible that the solution for $u_n$ oscillates
about a straight line of slope $-1$ (data for $u_0=0.4$ and $u_0=-1.0$).
For certain discrete set of initial values, the value of $u_n, n=
1,2,3,\dots$ etc.\ becomes exactly zero (close to data for
$u_0=0.9995$).  Such initial values are singular as the recurrence
relation becomes ill defined. However, even generic initial starting
points $0<u(0)<1$ are also pathological since eventually the
recurrence relation drives $u_n<0$ for $n$ sufficiently large, as we
will prove in the next section.  This is unphysical since $u_n$
represents a positive semi-definite operator.

\section{Asymptotic solutions}\label{section5}

Let us now discuss some of the properties of the solutions of the
equation~(\ref{recurrence}).  First, it is useful to derive the
``integral form'' of this expression. One can write
\begin{equation}
u_{k+1}-u_k=u_k-u_{k-1}+g^2\left(1-\frac{(k+1)^2}{u_k^2}\right).
\label{eq_a1}
\end{equation}
Let us then sum left and right sides of this equation for $k$ running
from zero to some integer $n$.  Two terms on the left as well as the
first two terms on the right are the so-called ``telescoping sum'' and
can easily be summed up to yield
\begin{equation}
u_{n+1}-u_0=u_n+g^2\sum_{k=0}^{k=n}\left(1-\frac{(k+1)^2}{u_k^2}\right).
\label{eq_a2}
\end{equation}
Moving $u_0$ to the right hand side and $u_n$ to the left, one can
repeat the summation again for $n$ going from $0$ to some other
integer $(l-1)$, producing the following formula for generic $u_l$
\begin{equation}
u_l=u_0(l+1)+g^2 \frac{l(l+1)}{2}
-g^2\sum_{n=0}^{n=l-1}\sum_{k=0}^{k=n}
\left(\frac{(k+1)^2}{u_k^2}\right).
\label{eqfinal}
\end{equation}
The whole sequence of $u_l$'s is therefore determined by its first
term, $u_0$ (and $\lambda^2$ of course). Since we are interested in the
solution where $u_l>0$ for all $l>=0$, we will therefore consider what
happens only for $u_0>0$ as well.  One special solution of this
recursion relation is easy to find, i.e.\
\begin{equation}
{\rm if} \qquad u_0=1\qquad \forall\,g\qquad {\rm then}\qquad
u_l=(l+1)\,.
\end{equation}  
\looseness=1 For the case of $u_0>1$ by induction one can prove that $u_l >
u_0(l+1)$. This follows immediately from the form of the
equation~(\ref{eq_a2}) since each term in the sum on the right hand
side is positive. However, one can improve this \pagebreak[3] estimate by using this
fact and~(\ref{eqfinal}), i.e.\
\begin{eqnarray}
u_l&=&u_0(l+1)+g^2 \frac{l(l+1)}{2}
-g^2\sum_{n=0}^{n=l-1}\sum_{k=0}^{k=n}
\left(\frac{(k+1)^2}{u_k^2}\right)
\nonumber\\
&>&u_0(l+1)+g^2 \frac{l(l+1)}{2}
-g^2\sum_{n=0}^{n=l-1}\sum_{k=0}^{k=n}
\left(\frac{1}{u_0^2}\right)
\nonumber\\
&>&u_0(l+1)+g^2\left(1-\frac{1}{u_0^2} \right) \frac{l(l+1)}{2}\,.
\end{eqnarray}
Since $u_0>1$ the coefficient at the $g^2$ term of this formula is
positive and so the $u_l$ is growing at least quadratically with $l$.
One can improve this estimate once more by noticing that since
eventually $u_l\,\sim\,l^2$ one can iterate~(\ref{eqfinal}) one more
time.  The actual form of the asymptotic expression for $u_l$ can be
found from the self-consistency condition and~is
\begin{equation}
u_l=g^2\frac{l(l+1)}{2}+ \alpha\,l +\beta\,{\rm ln}(l)+\gamma + {o}(l)\,,
\end{equation}
where $\alpha,\beta$ and $\gamma$ are some functions of $\lambda$ and
$u_0$.

Let us now turn to a case of $u_0<1$. Analysis of the previous
paragraph has to be modified by essentially changing all $''>''$ signs
to the $''<''$ ones with one extra condition that $u_l>0$ for all
``previous'' $l$'s.  Indeed, assuming that $u_k >0\ \ \forall\,k < l$
one can use induction to prove (using~(\ref{eq_a2}) as before) that
\begin{equation}
0<u_l <u_0(l+1)\,.
\end{equation}
Notice that the condition that all the previous $u_l$'s are positive
is essential, since it is the absolute value that matters in the
nonlinear part of~(\ref{eq_a2}) (i.e.\ $1/u_l^2$ ).  Let us assume
that up to a particular $l$ this condition is satisfied.  Then, using
equation~(\ref{eqfinal}) one has that
\begin{eqnarray}
u_l&=&u_0(l+1)+g^2 \frac{l(l+1)}{2}
-g^2\sum_{n=0}^{n=l-1}\sum_{k=0}^{k=n}
\left(\frac{(k+1)^2}{u_k^2}\right)
\nonumber\\
&<&u_0(l+1)+g^2 \frac{l(l+1)}{2}
-g^2\sum_{n=0}^{n=l-1}\sum_{k=0}^{k=n}
\left(\frac{1}{u_0^2}\right)
\nonumber\\
&<&u_0(l+1)+g^2\left(1-\frac{1}{u_0^2} \right) \frac{l(l+1)}{2}\,.
\end{eqnarray}
Here the coefficient at the quadratic part is now negative, since
$|u_0|<1$. This tells us that for sufficiently large $l$ one clearly
has that
\begin{equation}
u_l<u_0(l+1)+g^2\left(1-\frac{1}{u_0^2}
\right)\frac{l(l+1)}{2} <0 \qquad  {\rm for} \qquad
l>\frac{2u_0^3}{g^2(1-u_0^2)}
\end{equation}
which clearly contradicts our assumptions.  Therefore, for any
$0<u_0<1$ the recurrence relation~(\ref{recurrence}) will necessarily
produce at least one negative $u_l$.

In view that $u$'s represent a norm for the Hilbert space states and
have to be positive, it would seem then that $u_0<1$ is not a good
starting point. However, should the total size $n$ of the Hilbert
space be restricted~\cite{Polychronakos:2001mi} this is not so. \pagebreak[3] Since
the number operator $N$ eigenvalues roughly correspond to the radius
squared in the plane, this limiting value would correspond to a finite
size droplet. It quite possible that some of the $u_0$'s which are
less then one would be allowed.  We are currently investigating this
possibility.

\acknowledgments

We would like to thank C.~Burgess, B.~Dolan, A.~Khare, R.B.~MacKenzie
and V.P.~Nair for useful discussions and NSERC of Canada for financial
support.

\end{document}